\documentclass[twocolumn,amssymb,amsmath,aps,prb,showpacs,floatfix,superscriptaddress]{revtex4}

\usepackage{graphicx}

\begin{document}

\title{Broadband dielectric response of CaCu$_3$Ti$_4$O$_{12}$: From dc to the electronic transition regime}

\author{Ch.~Kant}
\author{T.~Rudolf}
\author{F.~Mayr}
\author{S.~Krohns}
\author{P.~Lunkenheimer}
\affiliation{Experimental Physics~V, Center
for Electronic Correlations and Magnetism, University of Augsburg,
86135~Augsburg, Germany}

\author{S.~G.~Ebbinghaus}
\affiliation{Solid State Chemistry, University of Augsburg,
86135~Augsburg, Germany}

\author{A.~Loidl}
\affiliation{Experimental Physics~V, Center for Electronic
Correlations and Magnetism, University of Augsburg, 86135~Augsburg,
Germany}

\date{\today}

\begin{abstract}
We report on phonon properties and electronic transitions in
CaCu$_3$Ti$_4$O$_{12}$, a material which reveals a colossal
dielectric constant at room temperature without any ferroelectric
transition. The results of far- and mid-infrared measurements are
compared to those obtained by broadband dielectric and
millimeter-wave spectroscopy on the same single crystal. The
unusual temperature dependence of phonon eigenfrequencies,
dampings and ionic plasma frequencies of low lying phonon modes
are analyzed and discussed in detail. Electronic excitations below
4~eV are identified as transitions between full and empty
hybridized oxygen-copper bands and between oxygen-copper and
unoccupied Ti~3$d$ bands. The unusually small band gap determined
from the dc-conductivity $(\sim200$~meV) compares well with the
optical results.
\end{abstract}

\pacs{63.20.-e, 78.30.-j, 77.22.Ch}

\maketitle

\section{Introduction}

After first reports of very large dielectric constants of order up
to $10^5$ in CaCu${}_3$Ti${}_4$O${}_{12}$ (CCTO) in ceramic
samples, \cite{subraman00} single crystals, \cite{homes01} and
thin films, \cite{si02} experimental evidence has been provided
that these colossal values have their origin in Maxwell-Wagner
like relaxation phenomena, characteristic for inhomogeneous media.
Planar defects in single crystals or grain boundaries in ceramics
\cite{sinclair02} as well as contact phenomena and surface effects
\cite{lunkenhe02,lunkenhe04,krohns07} were considered as possible
sources for the unconventional dielectric response in CCTO.
However, also intense search for intrinsic mechanisms still keeps
going on, an example being a recent report on nanoscale Ca/Cu
disorder.\cite{zhu07}

In addition to this unsettled dispute about the origin of colossal
dielectric constants in CCTO, another interesting phenomenon was
detected: The dielectric constant as measured by far-infrared
(FIR) spectroscopy is as large as 80 at room temperature and
increases with decreasing temperature \cite{homes01,homes03}
contrary to what is expected for a normal anharmonic solid. This
effect was investigated in some detail by Homes \textit{et
al}.\cite{homes03} and explained in terms of charge-transfer
processes.

The present investigation deals with the following topics: i) The
complete phononic response, which has been measured by FIR
spectroscopy as function of temperature, is analyzed in full
detail: eigenfrequencies, dampings, and ionic plasma frequencies
are determined for all modes to study the unusual temperature
dependence. Our results are compared to published results
\cite{homes01,homes03} and to model calculations of the lattice
dielectric response of CCTO from first
principles.\cite{he02,he03,mcguinne05} ii) Within the low
frequency reflectivity spectrum, which is dominated by phonon
modes, we detect an unusually large number of crossing points in
the reflectivity ($\partial R /\partial T = 0 $), which seems to
be too significant to be ignored or to be explained by accidental
effects. iii) Electronic excitations for energy transfers up to
4~eV are studied via the dynamic conductivity and are compared to
ab-initio band structure calculations.\cite{he02} iv) The dynamic
conductivity and dielectric constant from infrared and
millimeter-wave spectroscopy are directly compared to broadband
dielectric results to visualize the full dielectric response of
CCTO to electromagnetic fields over 15 decades in frequency and
finally, v), we derive the band gap from the dc conductivity
obtained from the dielectric results and find good agreement with
the theoretically predicted optical band gap, which is in accord
with the IR results.

\section{Experimental details}

Single crystals were grown by the floating-zone technique using a
growth furnace equipped with two 1000~W halogen lamps, with the
radiation focused by gold-coated ellipsoidal
mirrors.\cite{krohns07} Polycrystalline bars, prepared as reported
in Ref.~\onlinecite{lunkenhe04}, cold-pressed and sintered in air
for 12~h at 1000$^\circ$C, served as seed and feed rods. The rods
were rotated with a speed of 30~rpm, while the feed was kept
still. The growth rate was adjusted to 5~mm/h. Crystal growth was
performed in oxygen (flow rate 0.2~l/min) at a pressure of 4~bar,
to avoid thermal reduction of copper. High purity single crystals
with a lattice constant of 0.7391~nm and free of impurity phases
were obtained. CaCu$_3$Ti$_4$O$_{12}$ belongs to a unique class of
perovskite derived structures in which the TiO$_6$ octahedra are
strongly tilted to form an ideal square planar coordination for
the Cu cations.\cite{bochu79} The tilting of the octahedra and the
concomitant non-cubic site symmetry of Ti$^{4+}$ strongly reduces
the possibilities for off-center displacements and rules out
long-range polar order of the Ti ions. Thus this class of
compounds usually does not display
ferroelectricity.\cite{subraman00}

For the dielectric measurements silver paint contacts were applied
to opposite faces of the disc like single crystals. The complex
dielectric permittivity as function of temperature was measured
over nine frequency decades from 1~Hz up to 1.3~GHz (for
experimental details see Ref.~\onlinecite{boehmer89}). Additional
measurements using a quasioptic spectrometer in Mach-Zehnder
configuration were performed in a frequency range from 60 to
120~GHz.\cite{gorshuno05} In the far- and mid-infrared range,
reflectivity measurements were carried out using the Bruker
Fourier-transform spectrometers IFS 113v and IFS 66v/S, which both
are equipped with He bath cryostats. In most cases the
reflectivity spectra were directly analyzed using a generalized
oscillator model with four parameters per phonon
mode.\cite{gervais83,kuzmenko} To calculate the dielectric loss
from the experimentally obtained reflectivity, we used a smooth
$\omega^{-h}$ extrapolation to high frequencies. The low-frequency
extrapolation was based on the measured dielectric data.

\section{Results and Analysis}

\begin{figure}[tbp]
  \includegraphics[width=8cm,clip]{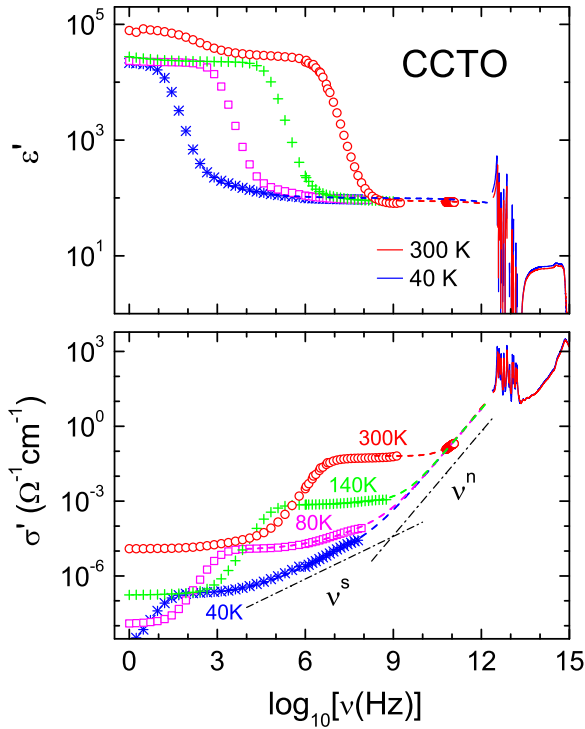}
  \caption{\label{fig1} Dielectric constant and dynamic
  conductivity of CCTO over 15 decades in frequency for various temperatures. Symbols and the lines
  at $\nu > 10^{12}$~Hz show experimental data. Dashed lines are fits of the data beyond the relaxation taking into
  account dc and ac conductivities, the latter described by sub-
  and super-linear power laws $\nu^s$ and $\nu^n$, respectively (dash-dotted lines, see text).}
\end{figure}

A survey of the dielectric constant $\varepsilon'(\omega)$ and the
dynamic conductivity $\sigma'({\omega})$ over 15 decades in
frequency is shown in Fig.~\ref{fig1} for temperatures between
40~K and 300~K. Fig.~\ref{fig1} impressively documents how, at
least at room temperature, the ``relaxational'' dielectric
response for frequencies below 10~GHz is decoupled from the ionic
($\nu = 10^{12}$~Hz - $2\times10^{13}$~Hz) and from the electronic
($\nu > 2\times10^{13}$~Hz) processes. The upper frame indicates
that the relatively high intrinsic dielectric constant of the
order of 100 detected at the higher frequencies and lower
temperatures of the dielectric
experiments\cite{subraman00,homes01,sinclair02,lunkenhe02,krohns07}
can be ascribed to the ionic polarizability. $\varepsilon_\infty$,
defined as $\varepsilon'$ beyond the phonon modes and determined
by the electronic polarizability only, is well below 10. The
intrinsic  conductivity of CCTO in the lower frame of
Fig.~\ref{fig1}, specifically at 40~K, follows a universal
behavior where the dc conductivity with $\nu^0$ at low frequencies
is followed by Jonscher's universal dielectric response with
$\nu^{0.57}$(Ref.~\onlinecite{jonscher77}) and by a super linear
power-law with $\nu^{1.39}$ at even higher frequencies. Fits using
this approach are indicated as dashed lines for $\varepsilon'$ and
$\sigma'$ in Fig. 1. This sequence of dc, as well as sub-linear
and super-linear ac conductivity regimes has been observed in a
number of disordered semiconductors and transition-metal
oxides.\cite{lunkenhe03a}

\subsection{Phonon excitations}

\begin{figure}[tbp]
  \includegraphics[width=8cm,clip]{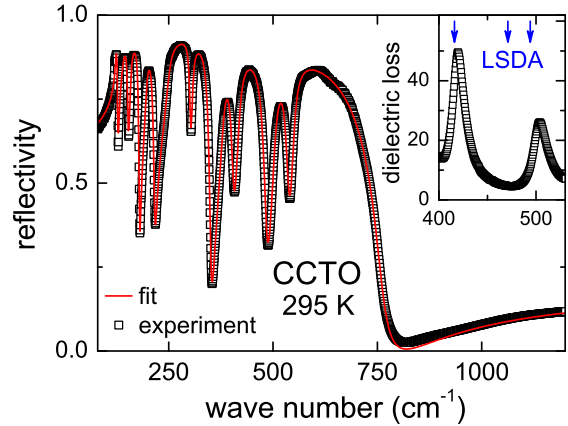}
  \caption{\label{fig2}(Color online) Reflectivity spectrum of CCTO at 295~K.
    The solid line is the result of a fit using ten oscillators as
    described in the text. Inset: Dielectric loss obtained by Kramers-Kronig
    analysis. The arrows indicate the positions of expected loss peaks
    according to Ref.~\onlinecite{he03}. Note that the predicted resonance at
    471~cm$^{-1}$ is completely absent.}
\end{figure}

Fig.~\ref{fig2} shows the measured reflectivity of CCTO at 295~K
including the results of a fit using ten oscillators described by
four parameters each.\cite{gervais83} The theoretical modeling
which has been used in this work is outlined in detail in Ref.
\onlinecite{rudolf07}. For each phonon mode the fit parameters are
the transverse optical (TO) and longitudinal optical (LO)
eigenfrequencies $\omega_{TO}$ and $\omega_{LO}$ and the damping
functions $\gamma_{TO}$ and $\gamma_{LO}$. In addition, the
electronic polarizability is taken into account by
$\varepsilon_\infty$, which is treated as free parameter. This
model yields an almost perfect description of all eigenmodes as
observed in the reflectivity spectrum. Similar results have been
obtained for a series of measurements at different temperatures
down to 5~K. From fits up to 2000~cm$^{-1}$ $\varepsilon_\infty$,
which is due to electronic polarizability only, has been
determined. $\varepsilon_\infty$ obtained from these fits was
found to scatter between 6 and 7, with no systematic temperature
variation. Hence, for all further analysis we used an average
value of 6.5 for all temperatures. A list of the LO and TO mode
frequencies and dampings as well as the effective ionic plasma
frequencies $\Omega$ and the dielectric strengths $\Delta
\varepsilon$ are given in Tab.~{\ref{tab1}}. For the definition of
plasma frequency and dielectric strength see
Ref.~\onlinecite{rudolf07}. The experimental results documented in
Tab.~\ref{tab1} compare reasonably well with first-principle
calculations of the TO eigenfrequencies of CCTO within local
spin-density approximation (LSDA) by He \textit{et
al}.,\cite{he03} which are shown in the second row of
Tab.~\ref{tab1}. All experimentally observed eigenfrequencies lie
in a frequency range of approximately $\pm$10~cm$^{-1}$ when
compared to the theoretical predictions. However a mode of
moderate strength, predicted to occur at 471~cm$^{-1}$
(Ref.~\onlinecite{he03}) is fully missing in the reflectivity
data. Indeed while the symmetry of the crystal allows for 11
IR-active modes, only ten are observed experimentally. A closer
look into Fig.~\ref{fig2} shows that fit and experimental result
almost coincide and it is hard to believe that an extra mode of
considerable strength can be hidden in this reflectivity spectrum,
if not two eigenfrequencies are accidentally degenerated within
$\pm$5~cm$^{-1}$. To check this possibility in more detail, the
inset of Fig.~\ref{fig2} shows the dielectric loss vs. wave number
in the frequency range from 400 to 530~cm$^{-1}$. The arrows in
the inset indicate the eigenfrequencies as theoretically
predicted. While theory meets the modes close to 420 and
500~cm$^{-1}$ there is not the slightest indication of an
additional mode close to 471~cm$^{-1}$. Due to the rather low
intensity of the mode close to 500~cm$^{-1}$, it seems unreliable
that the missing mode is hidden underneath it. This obvious
disagreement between theory and experiment remains to be settled.

\begin{table}[tbp]
\begin{ruledtabular}
\begin{tabular}{cccccccc}
  mode & \multicolumn{3}{c}{$\omega$ (cm$^{-1}$)} & \multicolumn{2}{c}{$\gamma$
(cm$^{-1}$)}&
  $\Omega$ (cm$^{-1}$) & $\Delta \varepsilon$ \\
  5~K&LSDA\cite{he03}&TO&LO&TO&LO&&\\
  \hline
   1 & 125 & 119.2 & 129.8 & 11.8 & 1.3 & 478.2 & 16.1 \\
   2 & 135 & 134.5 & 152.7 &  5.2 & 2.5 & 591.1 & 19.4 \\
   3 & 158 & 158.1 & 181.1 &  4.2 & 2.4 & 630.7 & 15.9 \\
   4 & 199 & 195.1 & 216.1 &  5.2 & 4.7 & 599.3 &  9.4 \\
   5 & 261 & 250.4 & 303.9 & 11.9 & 3.4 & 914.6 & 13.3 \\
   6 & 310 & 307.7 & 352.9 &  4.5 & 4.3 & 799.9 &  6.8 \\
   7 & 385 & 382.9 & 407.7 &  8.8 & 5.5 & 608.7 &  2.5 \\
   8 & 416 & 421.3 & 487.3 &  9.1 &12.2 & 920.6 &  4.8 \\
   9 & 471 \\
  10 & 494 & 506.9 & 542.4 & 13.2 &10.2 & 677.9 &  1.8 \\
  11 & 547 & 551.6 & 760.3 & 10.2 &32.2 &1333.8 &  5.8 \\
\end{tabular}
\end{ruledtabular}
\caption{\label{tab1}Eigenfrequencies $\omega$ and
  dampings $\gamma$ of both transverse (TO) and longitudinal optical
  (LO) phonon modes of CCTO at 5~K. Additionally the plasma
  frequencies  $\Omega$  and dielectric strengths $\Delta \varepsilon$ are provided for each mode.
  The second row contains the frequencies of IR-active
  phonon modes calculated by He \textit{et al}..\cite{he03}}
\end{table}

\begin{figure}[tbp]
  \includegraphics[width=8cm,clip]{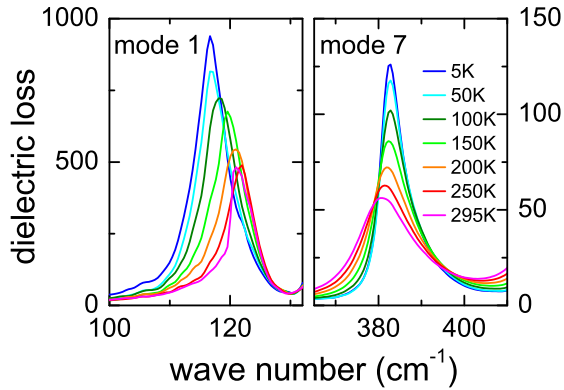}
  \caption{\label{fig3} Dielectric loss of mode 1 (left panel) and mode 7 (right panel) as function of wave number for various temperatures.}
\end{figure}

In the following we will address the different phonon modes with
numbers as indicated in Tab.~\ref{tab1}. A closer inspection of
this table reveals rather unusual details. Specifically, for modes
1 and 5, $\gamma_{TO}$ is much larger compared to $\gamma_{LO}$,
contrary to what is expected in a canonical anharmonic solid. This
may be at least partly related to the unconventional temperature
dependence of eigenfrequencies and damping (see below). In what
follows we will give a detailed description of the temperature
dependence of some of the polar phonon modes in CCTO. The modes
can be grouped into two fractions: the first 5 modes reveal an
unusual temperature dependence which cannot be explained by normal
anharmonic effects.\cite{cowley63} Phonons number six up to number
eleven can be classified as phonon excitations of a classical
anharmonic solid. As prototypical examples, Fig.~\ref{fig3} shows
the dielectric loss of phonon 1 and phonon 7 as function of wave
number for a series of temperatures. With decreasing temperature,
phonon 1 which lies close to 120~cm$^{-1}$ at room temperature,
broadens, shifts to lower frequencies and strongly increases in
dielectric strength. On the contrary phonon 7, which appears close
to 380~cm$^{-1}$ at 295~K, becomes narrow and shifts to higher
frequencies on lowering the temperature, a behavior reflecting
anharmonicity due to phonon-phonon scattering processes. As will
be shown later, in this case the dielectric strength almost
remains constant which is expected in a purely ionic solid with no
charge transfer processes and no ferroelectric instability.

\begin{figure}[tbp]
  \includegraphics[width=8cm,clip]{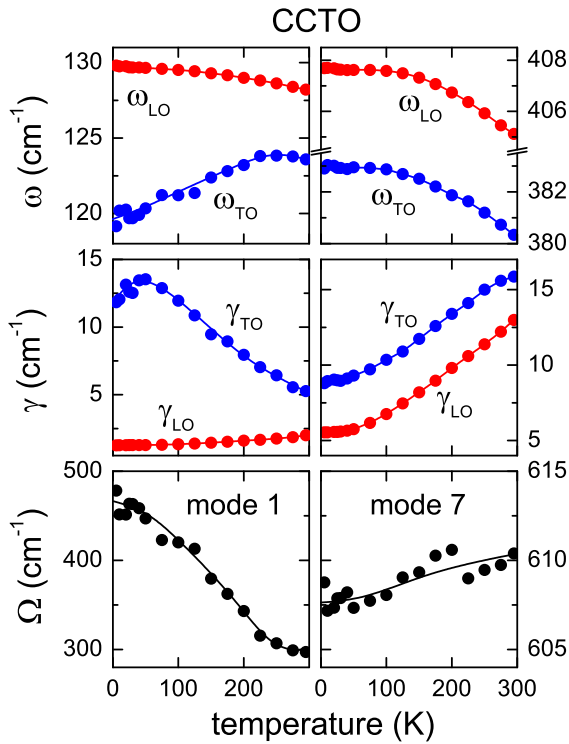}
  \caption{\label{fig4}(Color online) Eigenfrequencies, dampings and
  plasma frequencies for mode 1 (left frames) and mode 7 (right frames)
  as function of temperature. The lines are drawn to guide the eyes.}
\end{figure}

The results of a detailed 4-parameter analysis are documented in
Fig.~\ref{fig4}: LO and TO eigenfrequencies (upper frames) and
dampings (middle frames), as well as the ionic plasma frequencies
(lower frames) are shown for phonon 1 (left frames) and phonon 7
(right frames). For mode 1 the TO mode softens considerably and
its damping is unusually large and increases on decreasing
temperature. As has been documented already by Homes \textit{et
al}.,\cite{homes01,homes03} the plasma frequency increases by as
much as 60\% when the temperature is lowered from room temperature
down to 5~K. On the other hand, mode 7 shows conventional
behavior. LO and TO eigenfrequencies slightly increase and the
inverse life times decrease when temperature is lowered. The
plasma frequency almost remains constant at a value of (609 $\pm$
2)~cm$^{-1}$, which certainly is within the experimental
uncertainties. As the ionic plasma frequency, which corresponds to
the effective charges, is proportional to the difference of the
squared LO and TO eigenfrequencies, it is obvious that the
increase of the plasma frequency $\Omega$ of mode 1 predominantly
corresponds to the softening of the transverse optic mode (see
upper left frame in Fig.~\ref{fig4}). At present it is unclear
whether this observation indicates an underlying ferroelectric
instability or points towards charge transfer processes as driving
forces.

\begin{figure}[tbp]
  \includegraphics[width=7.5cm,clip]{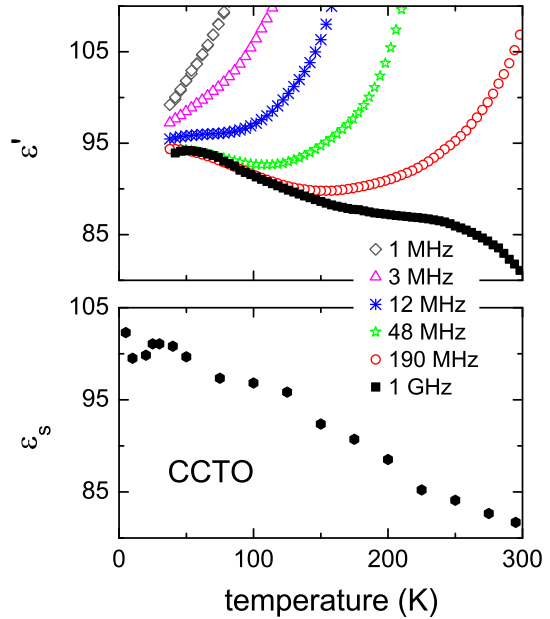}
  \caption{\label{fig5}(Color online) Upper panel: Dielectric
  constant for various frequencies as function of temperature
  obtained by dielectric spectroscopy. Lower panel: static dielectric
  constant derived from FIR experiments.}
\end{figure}

As documented in Tab.~\ref{tab1} for 5~K, we calculated the
dielectric strength for all modes and determined these values as
function of temperature. The lower frame in Fig.~\ref{fig5} shows
the static dielectric constant $\varepsilon_s$, which corresponds
to $\varepsilon_\infty$ plus the sum over the dielectric strengths
of all modes. According to this FIR result, $\varepsilon_s$
increases from 83 at room temperature to approximately 100 at low
temperatures. This has to be compared to measurements of the
dielectric constant at GHz frequencies, which corresponds to the
intrinsic static dielectric constant. These results are documented
in the upper frame of Fig.~\ref{fig5}. The GHz dielectric constant
indeed roughly follows the FIR results, especially it shows
similar temperature dependence. The strong increase of
$\varepsilon'$ as detected at lower frequencies corresponds to
Maxwell-Wagner like effects.\cite{lunkenhe02,lunkenhe04,krohns07}

\begin{figure}[tbp]
  \includegraphics[width=8cm,clip]{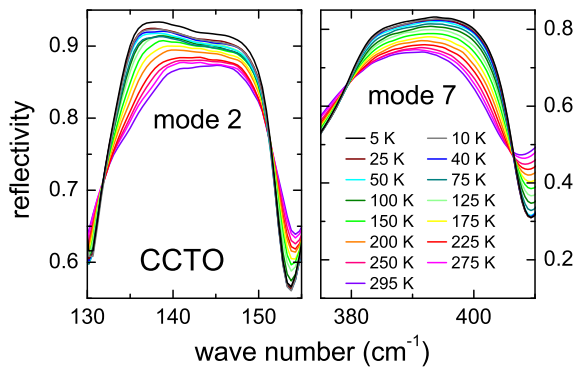}
  \caption{\label{fig6} Raw reflectivity data for mode 2
  (left panel) and mode 7 (right panel) for various temperatures. Note
  the isosbestic points, i.e. points of temperature independent reflectivity.}
\end{figure}

In discussing the phonon properties we would like to point towards
another interesting phenomenon, whose nature and origin are
unclear at present. Each pair of TO and LO modes creates a
rectangular shaped band in the reflectivity: In an ideal harmonic
solid one would expect that the reflectivity $R$ is close to unity
between the TO and the LO modes. The decrease of $R(\omega)$ with
increasing temperature follows from an increasing anharmonicity.
In the temperature dependent reflectivity spectra of CCTO, each
band exhibits two striking crossing points. In optical
spectroscopy of chemical species, a so-called isosbestic point
usually defines a point on the wavelength scale, where two species
have exactly the same absorption. Isosbestic points have sometimes
also been identified in the dynamical conductivity of transition
metal oxides and were explained in terms of spectral weight
transfer driven by strong electronic correlations.\cite{okimoto95}
Quite generally it can be stated that whenever a system can be
described by a superposition of two components with dynamic
quantities which only depend linearly on density, isosbestic
points are expected to occur.\cite{eckstein07} In particular this
also applies to the temperature dependence as long as the total
density is constant. In our case each reflectivity band exhibits
two wavelengths where the reflectivity exactly is temperature
independent, i.e. $\partial R/\partial T=0$. As an example
Fig.~\ref{fig6} shows the reflectivity at around 140 and
390~cm$^{-1}$. Close to each TO and LO mode we find these crossing
points where indeed the reflectivity is temperature independent
within experimental uncertainty. Again we would like to stress
that similar observations can be made for each reflectivity band
of CCTO. At present, however, it is unclear how the reflectivity
of CCTO can be described by the interaction of light with two
components. One could think of microscopic (electronic) phase
separation. Indeed reports on nanoscale disorder of Cu and Ca
sites from x-ray absorption fine structure
measurements\cite{zhu07} and reports of the coexistence of
strained and unstrained domains by scanning electron microscopy
\cite{fang05} provide some arguments in favor of this explanation.
One also could speculate that in CCTO the CuO$_2$ planes and the
TiO$_6$ octahedra behave like two independent components being
responsible for the occurrence of isosbestic points in all
absorption bands.

\subsection{Electronic exitations}

\begin{figure}[tbp]
  \includegraphics[width=8cm,clip]{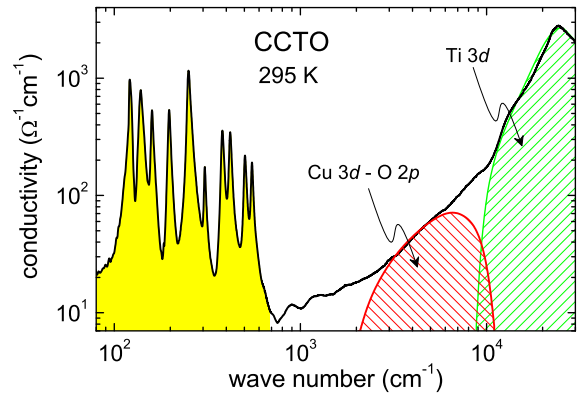}
  \caption{\label{fig7}(Color online) Dynamic conductivity of CCTO at
  295~K. The spectral response can be divided into three major contributions:
  a phonon part (shaded area), transitions into empty Cu~3$d$ - O~2$p$ orbitals
  (left hatched area) and transitions into unoccupied Ti~3$d$ states (right hatched area).}
\end{figure}

Finally we studied the low-lying electronic transitions up to
30000~cm$^{-1}$, corresponding to an energy of 3.8~eV. The results
are documented in Fig.~\ref{fig7}, which shows the real part of
the dynamic conductivity \textit{vs}. wave number on a
double-logarithmic scale. The spectral response below
700~cm$^{-1}$ is dominated by the phonon response. Beyond the
phonon regime the conductivity gradually increases up to
10000~cm$^{-1}$ and then shows a stronger increase with a peak
close to 24000~cm$^{-1}$ corresponding to 3.0~eV. We would like to
stress that the conductivity below 1~eV ($\approx
8000~\text{cm}^{-1}$) is relatively small (note the double
logarithmic scale of Fig.~\ref{fig7}) and slightly depends on the
extrapolation scheme beyond 4~eV, used for the Kramers-Kronig
analysis. However, all reasonable extrapolations yield similar
results with only slight differences in the tail towards the
lowest frequencies. When comparing our results to first principle
density-functional theory within the local spin-density
approximation (LSDA) from He \textit{et al}.,\cite{he02} we
identify the continuous increase of the conductivity from about
2000~cm$^{-1}$ to 10000~cm$^{-1}$ with transitions between the
empty and filled strongly hybridized Cu~3$d$ and O~2$p$ orbitals.
The filled bands are located just below the Fermi level while the
empty states extend up to 0.7~eV with a maximum close to 0.5~eV
and an onset at about 0.25~eV, which corresponds to approximately
2000~cm$^{-1}$. According to theory, the dominant peak close to
24000~cm$^{-1}$ ($\sim3.0$~eV), observed in Fig.~\ref{fig7}, can
be identified with transitions into predominantly $d$-derived
empty states from the Ti ions. In the LSDA calculations this
Ti~3$d$ band of mainly $t_{2g}$ character extends from 1.5 to
3.5~eV with a peak maximum close to 2.8~eV. Thus, overall the
dynamical conductivity at high frequencies could be determined by
the superposition of two electronic transition bands located at
around 0.75~eV and 3.0~eV as schematically indicated by the
hatched areas in Fig. \ref{fig7}. It should be noted that the
conductivity tail due to transitions between the hybridized
copper-oxygen bands extends to rather low frequencies. In the
chosen double-logarithmic plot, the band gap should be read off at
a limiting vertical decrease of $\sigma'(\nu)$ at low frequencies.
This is not observed in the data, most likely due to the mentioned
uncertainties at very low conductivity values or possible phonon
tails as expected for indirect transitions. Clearly, while our
results are not a proof of the bandstructure of CCTO, they at
least are compatible with the LSDA calculations, especially
concerning the predicted small band gap.\cite{he02}

\begin{figure}[tbp]
  \includegraphics[width=8cm,clip]{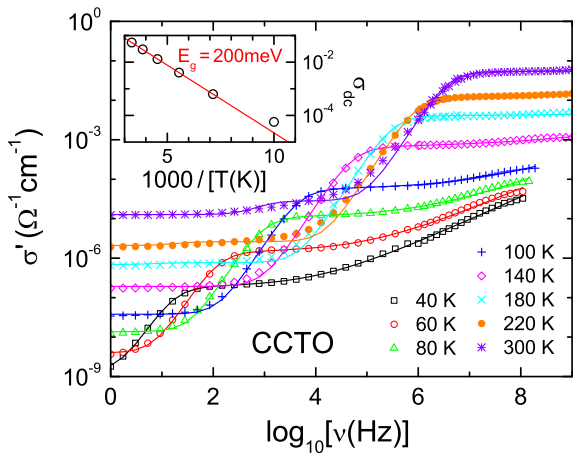}
  \caption{\label{fig8}(Color online) Dynamic conductivity over 9 decades
  of frequency for CCTO for various temperatures. Solid lines represent
  the results of an equivalent-circuit analysis (see text). The inset shows the resulting dc
  conductivity at $T\geq100$~K in an Arrhenius-type representation.}
\end{figure}

It can be expected, that this optical band gap also determines the
dc conductivity. In Fig.~\ref{fig8} we show the conductivity of
CCTO deduced from the dielectric experiments over 9 decades of
frequency for a series of temperatures.\cite{krohns07} At high
frequencies and low temperatures $\sigma'(\nu)$ of CCTO is
dominated by ac conductivity, resulting in a power-law increase
towards the highest frequencies. At the highest temperatures the
conductivity in the MHz to GHz range is purely of dc type and
independent of frequency. Towards lower frequencies, the
well-known Maxwell-Wagner relaxation leads to a step like decrease
of $\sigma'$. The dc plateaus for each temperature can easily be
identified. At 40~K the dc plateau is located between 50~Hz and
10~kHz and shifts to higher frequencies with increasing
temperature. It is located beyond 10~MHz for 300~K. These dc
conductivity plateaus also show up in the lower frame in
Fig.~\ref{fig1}.

The complex frequency-temperature dependence of the conductivity
can only be exactly analyzed utilizing an equivalent-circuit
analysis including elements for the bulk sample and surface
layers. The solid lines in Fig.~\ref{fig8} result from fits
assuming two $RC$ circuits for two types of barriers (e.g.,
external and internal) and one $RC$ circuit including ac
conductivity for the bulk sample. For details see Refs.
\onlinecite{lunkenhe02}, \onlinecite{lunkenhe04}, and
\onlinecite{krohns07}. The resulting dc conductivity at
$T\geq100$~K is indicated in the inset of Fig.~\ref{fig8} in an
Arrhenius type of presentation. The energy barrier derived from
the Arrhenius fit is of the order of 200~meV, in good agreement
with the theoretical band gap and consistent with the optical
results (cf. Fig.~\ref{fig7}). Here we assumed an intrinsic
semiconductor with constant mobility and a charge-carrier density
proportional to $\exp[-E_{g}/(2k_{B}T)]$. At lower temperatures,
deviations from Arrhenius behavior show up, which may be ascribed
to hopping conductivity of localized charge carriers as will be
discussed in a forthcoming paper. \cite{krohns07b}

\section{Concluding remarks}

In summary, our detailed optical characterization of CCTO and the
comparison with broadband dielectric spectroscopy performed on the
same single crystal revealed a number of unusual properties of this
material, in addition to the well-known colossal dielectric
constants.

We analyzed in detail the temperature dependence of the phonon
modes and determined LO and TO eigenfrequencies and dampings as
well as the ionic plasma frequencies. The low lying modes (numbers
1 to 5, see Table~\ref{tab1}) do not behave like phonons of normal
anharmonic solids. The TO modes soften and the plasma frequencies
strongly increase. At present it is unclear if this is due to an
underlying ferroelectric instability, which however does not lead
to a transition even for lowest temperatures or if this indicates
significant charge-transfer processes as has been assumed by Homes
\textit{et al}..\cite{homes03} The phonon modes which are higher
in frequency (numbers 6 to 11) exhibit canonical behavior; i.e.
eigenfrequencies and dampings reveal a temperature dependence
characteristic for an anharmonic solid, which is dominated by
phonon-phonon interactions. The damping of the TO mode at
120~cm$^{-1}$ shows a cusp close to the antiferromagnetic phase
transition. In CCTO the Cu$^{2+}$ 3$d$ electrons constitute almost
localized $S=1/2$ spins, which undergo Neel ordering close to
$T_{N}=24$~K.\cite{Kim02}  In strongly correlated electron systems
often strong spin-phonon coupling is observed.\cite{rudolf07} In
CCTO, however, only mode 1 shows an anomaly at $T_{N}$ and overall
it seems that the phonons are not strongly coupled to the spin
system.

The static dielectric constant arising from the phonon modes as
derived from the measurements of this work increases from roughly
80 at room temperature to approximately 100 at 5~K. It nicely
scales with the dielectric constants measured at 1~GHz by
dielectric spectroscopy. The purely electronic polarizability
leads to $\varepsilon_\infty = 6.5$.

As function of temperature, each reflectivity band exhibits
significant crossing points. At these ``isosbestic'' points the
reflectivity is completely independent of temperature. The
implications for eigenfrequencies, dampings, and strengths are
unclear. These crossing points are too significant and too well
defined to be explained by trivial effects of the temperature
dependence of the phonon modes. Isosbestic points are usually
explained as being due to two components with constant total
density. An identification of two components in CCTO is not
straightforward and these crossing points await a deeper
theoretical analysis.

Finally, we summed up the effective plasma frequencies for all
modes, which must correspond to the ratio of the squared effective
charges and the mass of all ions in the unit cell (see, e.g., Ref.
\onlinecite{rudolf07}). Experimentally we find a value of $\Omega
= 2500\text{~cm}^{-1}$, compared to the theoretical value of
2700~cm$^{-1}$ that assumes ideal ionicity for all atoms. In
calculating the ionic plasma frequency of CCTO, the main
contributions result from Ti$^{4+}$ and O$^{2-}$ while Ca$^{2+}$
and Cu$^{2+}$ ions contribute less than 1\%. The fact that the
experimentally observed plasma frequency is so close to that
calculated for an ideal ionic solid, demonstrates that at least
the TiO$_{6}$ octahedra are purely ionically bonded while the Cu-O
subsystem can reveal partly covalent bonds. This seems to be in
accord with the LSDA calculations. \cite{he02,he03}

The strong ionicity and the weak hybridization between the oxygen
2$p$ levels and the titanium 3$d$ states\cite{he02} indicate that
the underlying nature of the anomalies in CCTO are different to
the origin of ferroelectricity in perovskite oxides: These model
ferroelectrics require strong hybridization, charge distortion,
and covalency.\cite{cohen92} However also alternative routes to
ferroelectricity were proposed, taking into account the strong
polarizability of the O$^{2-}$ ion\cite{migoni76} and recently an
attempt has been made to explain the optical response of CCTO
utilizing these ideas.\cite{bussmann03}

The dynamic conductivity beyond the phonon modes is consistent
with two electronic excitations as theoretically predicted from
LSDA band structure calculations.\cite{he02} They can be ascribed
to transitions from the filled hybridized O~2$p$ and Cu~3$d$ bands
to the empty O~2$p$/Cu~3$d$ states and to the empty Ti~3$d$
orbitals, arsing close to 0.75~eV and 3.0~eV, respectively.

\begin{acknowledgments}
This research was supported by the European Commission via STREP:
NUOTO, NMP3-CT-2006-032644 and partly by the Collaborative
Research Program, SFB 484 (Augsburg). Stimulating discussions with
D. Vollhardt are gratefully acknowledged.
\end{acknowledgments}


\end{document}